# Anomalous Conductance Quantization in Carbon Nanotubes

M. J. Biercuk, N. Mason, J. Martin*, A. Yacoby*, C. M. Marcus

*Department of Physics, Harvard University, Cambridge, Massachusetts 02138, USA*

Conductance measurements of carbon nanotubes containing gated local depletion regions exhibit plateaus as a function of gate voltage, spaced by approximately $1e^2/h$, the quantum of conductance for a single (non-degenerate) mode. Plateau structure is investigated as a function of bias voltage, temperature, and magnetic field. We speculate on the origin of this surprising quantization, which appears to lack band and spin degeneracy.

Carbon nanotubes free of disorder are expected to behave as ideal quantum wires, with electrical conduction occurring through one-dimensional (1D) modes [1], each with conductance quantized in units of $e^2/h$. In a variety of physical systems, including gate-defined quantum point contacts [2] and cleaved-edge wires [3], such 1D behavior appears as conductance plateaus as a function of voltage on nearby gates, which act to reduce density in all or some of the wire and hence depopulate the 1D subbands. In gated heterostructure quantum point contacts, conductance steps of $2e^2/h$ are observed at zero magnetic field, the factor of two reflecting spin degeneracy of the subbands [4, 5]. By analogy, one would expect nanotubes to show either a single step of $4e^2/h$ —the four reflecting the four modes per subband associated with spin and band degeneracy [6, 7] — or two steps spaced by $2e^2/h$ if band degeneracy were lifted, for instance by strain [7]. In this Letter we report conductance plateaus in gated nanotube devices in various configurations, revealing an unexpected plateau spacing of ~ $1e^2/h$ at zero applied magnetic field.

Conductance quantization has previously been observed in multiwalled carbon nanotubes by immersing one end in a liquid conductor [8]. This study found quantization principally in units of $2e^2/h$, with additional plateaus appearing near $e^2/h$ under certain conditions. In single-walled nanotubes, multiple steps in dc current were reported in devices with highly resistive metal contacts [9, 10], and were attributed to populating higher 1D subbands.

The nanotubes used in this study were grown by chemical vapor deposition (CVD) from Fe catalyst on doped Si wafers (which serve as backgates) with 1μm thermal oxide ($SiO_2$), and contacted with ~ 15 nm of Pd [11] following growth. All measured devices had nanotube diameters in the range ~ 1.5 – 5 nm (actual diameters noted in figure captions) [12]. In some cases, the nanotubes were pushed with the tip of an atomic force microscope (AFM) to form a bend [13-15]. Three device configurations were investigated: bent tubes with side-gates (Fig 1b, inset), as well as unbent and bent tubes with local electrostatic top-gates (Fig. 3a, inset). It has previously been shown that bends create gate-depletable regions [14, 15], and that local gates affect only proximal sections of the tube [16]. The top-gated devices were made by depositing CVD-grown $SiO_2$ on the Pd-contacted nanotubes and patterning Cr/Au gates over the nanotubes using electron-beam lithography [16, 17]. Two-terminal differential conductance, $G = dI/dV$, was measured as a function of source-drain bias, $V_{SD}$, by applying dc+ac voltage, $V = V_{SD} + V_{ac}$ (with $V_{ac}$ in the range 50-180 μV), and separately measuring dc and ac currents. Fourteen devices in these configurations showed qualitatively similar behavior.

Figure 1a shows characteristic plateau features in differential conductance for an intentionally bent tube with a local side gate near the bend (device shown in Fig. 1b) as a function of source-drain voltage, $V_{SD}$, and side-gate voltage, $V_{SG}$, with $V_{SG}$ held fixed for each trace. Plateaus appear as bunched traces, i.e., places where the conductance changes little as $V_{SG}$ is changed. These plateaus are also apparent in slices taken at fixed source-drain voltage (colored vertical lines in Fig. 1a) as a function of side-gate voltage, as shown in Fig. 1b. In this device, high-bias differential conductance saturates at ~ $3.3e^2/h$, somewhat below the ideal value of $4e^2/h$, presumably due to backscattering at the contacts or within the tube. To account for this, a series resistance, $R_S$, is subtracted to bring the high-field saturation to ~ $4e^2/h$. Several plateaus are visible, at both low and high bias. Note that plateaus around zero bias as well as at higher bias ($V_{SD}$ ~ 30 mV) are spaced by roughly $e^2/h$. Plateaus around zero bias show considerable overshoot, while those at high bias are typically much flatter. Smooth evolution between these two sets of plateaus with changing $V_{SD}$ is also evident, and revealed more clearly in Fig. 2. Conductance shows a dip around zero bias that deepens as temperature is lowered. Further, at temperatures below ~ 15 K single-electron charging is evident when the conductance is below the first plateau (Fig. 1c), while in the high conductance region an oscillatory pattern is observed near $V_{SD} = 0$ (Fig. 1d), which we associate with Fabry-Perot interference [18]. This structure may affect the positions of the plateaus, especially near $V_{SD} = 0$.

The transconductance, $dG/dV_{SG}$, for the same device (inset Fig. 1b) is shown in Fig. 2a. Transconductance highlights transitions between plateaus, with dark regions representing the plateaus and bright regions representing transitions. Figure 2a shows that as $|V_{SD}|$ is increased each transition splits into two. At larger bias, these split transitions cross, restoring the original number of plateaus. (It is near this re-crossing value of bias, $V_{SD}$ ~ 30 mV, that the high-bias cut in Fig. 1b is taken.) The resulting diamond pattern of splitting and rejoining transitions can be interpreted in the context of transport through quantized modes: If one assumes that each transition corresponds to the entering of a 1D mode into the transport window defined by the source and drain Fermi energies, then the diamond pattern in transconductance follows the evolution of mode energies with $V_{SD}$ and $V_{SG}$. This is the standard non-interacting picture of nonlinear "half-plateaus" in quantum point contacts [19, 20].

Experimental transconductance data (Fig. 2a) can be compared to various schemes for the evolution of 1D modes. In the simplest picture of four conduction modes with spin and band degeneracy [21], one expects a single transition from $G = 0$ to $G = 4e^2/h$ around zero bias as $V_{SG}$ is increased, and a single half-plateau with $G = 2e^2/h$ at high bias (Fig. 2b). With one degeneracy lifted (e.g., band degeneracy lifted by strain), the simple picture gives transconductance features as in Fig. 2c, with plateaus and transitions spaced by $2e^2/h$ around zero bias and half-plateaus at $e^2/h$ and $3e^2/h$. With all degeneracies lifted, this picture gives four plateaus and four transitions each spaced by $e^2/h$, and half-plateaus at 1/2, 3/2, 5/2 and 7/2 times $e^2/h$ (Fig. 2d). Surprisingly, the experimental data most resemble the schematic in Fig. 2d.

Within an interpretation of separate 1D modes, positions of the half-plateaus from Fig. 2a give a value for the 1D-mode energy spacing of $\Delta_{1D} \sim 55$ meV between the first and second plateau, with a spacing in gate voltage of ~ 1.8 V. Together, these give a coupling efficiency $\alpha \equiv \delta E / e\delta V_{SG} \sim 0.03$, which describes how far the Fermi level of the nanotube shifts with applied side gate voltage [22]. The coefficient $\alpha = C_L/(e^2 \partial n/\partial \mu)$, where $C_L^{-1} = (C_L^G)^{-1} + (e^2 \partial n/\partial \mu)^{-1}$, contains both a geometrical capacitance per unit length, $C_L^G$, and a term reflecting the kinetic energy (per unit length) required to increase the depth of the Fermi sea ($n$ is the density per unit length, $\mu$ is the chemical potential). Since $\alpha \ll 1$ (kinetic capacitance dominates geometric capacitance), we may approximate $\alpha \sim C_L^G/(e^2 \partial n/\partial \mu)$. The kinetic capacitance component can be calculated from a linear dispersion relation, $\delta E = \hbar v_F \delta k$ and $\delta k = \pi \delta n / M$ (where $M$ is the number of modes), giving $e^2 \partial n/\partial \mu = e^2/\pi \hbar v_F \sim 100$ pF/m per mode, where $v_F \sim 8 \times 10^5$ m/s is the typical Fermi velocity. The measured value $\alpha \sim 0.03$, extracted when only one mode is present, can then be used to calculate $C_L^G \sim 3$ pF/m.

The estimate of $C_L^G$ above has assumed fully lifted degeneracies. This interpretation is supported by a comparison of the length of the mechanical bend ($\sim 500$ nm) with the effective device length $L$ associated with both Coulomb blockade and Fabry-Perot oscillations, determined using $C_L^G$. The period of Coulomb blockade oscillations, $\Delta V_{SG} = e(LC_L^G)^{-1} \sim 0.05$ V, which corresponds to the addition of one electron to a length $L$, gives $L \sim 1$ $\mu m$. The period in $V_{SG}$ of the Fabry-Perot pattern corresponds to a change in wave vector by $\delta k_F = \pi/L$, hence a change in carrier density by $\delta n = 4\delta k_F/\pi$, taking all modes to be occupied near the highest plateau, where Fabry-Perot is measured. Relating the Fabry-Perot period $\Delta V_{SG} = 0.3$V to density, $e\delta n = C_L^G \Delta V_{SG}$, gives an independent estimate of effective device length, $L = 4e(\Delta V_{SG} C_L^G)^{-1} \sim 700$ nm, using $C_L^G$ from above.

We note that interpreting the measured half-plateau positions in terms of 1D subbands would require a nanotube diameter of $\sim 15$ nm, inconsistent with atomic force microscope measurements for all devices. If the band structure of the nanotube were modified substantially by the presence of a mechanical defect and a nonuniform gate, the 1D subband spacing could be reduced to the values we observe. Such an interpretation, however, would not explain the appearance of conductance steps with spacings of $e^2/h$.

In all measured devices, the differential conductance was everywhere less than $4e^2/h$. Subtracting a series resistance to bring the large-bias conductance to $4e^2/h$ typically yielded plateaus and half-plateaus spaced by $\sim e^2/h$ [23]. In no case could we find an appropriate series resistance that would give plateaus separated by $2e^2/h$ or $4e^2/h$.

Figure 3 demonstrates the effect of the gates in producing spatially localized depletion regions. Conductance of an unbent tube with two top gates shows plateau structure as a function of either gate. The two gates evidently act independently, each depleting different regions of the tube locally, leading to the square pattern seen in the inset of Fig. 3a. Two gates influencing the same region of the tube would produce a diagonal pattern instead.

Three of the measured devices showed a nonconductive region as a function of gate voltage with high conductance regions on either side, presumably reflecting an energy gap in a semiconducting nanotube [24]. Data for one such device, an intentionally bent tube without local gates, are shown in Fig. 3(b). Both the hole-like regime (at negative gate voltage), and the electron-like regime (at positive gate voltage) show $e^2/h$ conductance plateaus as a function of gate voltage. Figure 3(c) shows plateaus structure both around zero bias and higher bias, with spacings between plateaus of $\sim e^2/h$ in another bent-tube device as a function of top-gate voltage over the bend and source-drain voltage.

Figure 4 shows the low-bias conductance plateaus of the side-gated device (Fig. 1b, inset) as a function of temperature and magnetic field. No series resistance has been subtracted from these data. The plateau near $e^2/h$, which shows considerable overshoot at low temperature, rises to a value close to $1.5e^2/h$ with increasing temperature, while the subsequent plateaus are smoothed with increasing temperature, but do not rise significantly. Plateaus measured for this device at high bias show little change with increasing temperature. Magnetic field applied perpendicular to the tube axis has little effect on plateau structure (Fig. 4(b)). Here, the Zeeman splitting at $B = 8$ T, $g\mu_B B \sim 0.9$ meV, is greater than thermal energy ($k_B T \sim 0.15$ meV) but less than the voltage bias ($eV_{SD} = 2$ meV). Comparable ratios of magnetic field to bias energies induce significant change in $G(V_G)$ curves in GaAs quantum point contacts with spin degenerate levels.

Possibly relevant to our results is the appearance of a zero-magnetic-field conductance plateau near $e^2/h$ [25], or more commonly closer to $0.7(2e^2/h)$ [26-28], in gate-defined semiconductor quantum point contacts and wires. A theoretical model of this so-called 0.7 structure in quantum point contacts involving the formation of an ordered electronic state in 1D [29] may be relevant in nanotubes as well. Finally, we note that previous studies of Pd nanowires have observed conductance plateaus at $\sim e^2/h$, which the authors suggested may indicate ferromagnetism or near-ferromagnetism in Pd nanostructures [30] (though $\sim e^2/h$ plateaus were also seen in similar Ti structures). The plateau structure reported here is most evident in Pd contacted nanotubes, however we believe this is primarily due to the highly transparent contacts obtained with Pd rather than an effect of near-ferromagnetism in the leads, as magnetic field sweeps at both high field and around zero field showed no hysteresis or change in plateau structure.

This work was supported by NSF-NIRT (EIA-0210736), and ARO/ARDA (DAAD19-02-1-0039 and -0191) and Harvard CIMS. M.J.B. acknowledges support from the NSF and an ARO-QCGR Fellowship. N. M. acknowledges support from the Harvard Society of Fellows.

References:

*Permanent Address: Department of Physics, Weizmann Institute, Rehovot, Israel.

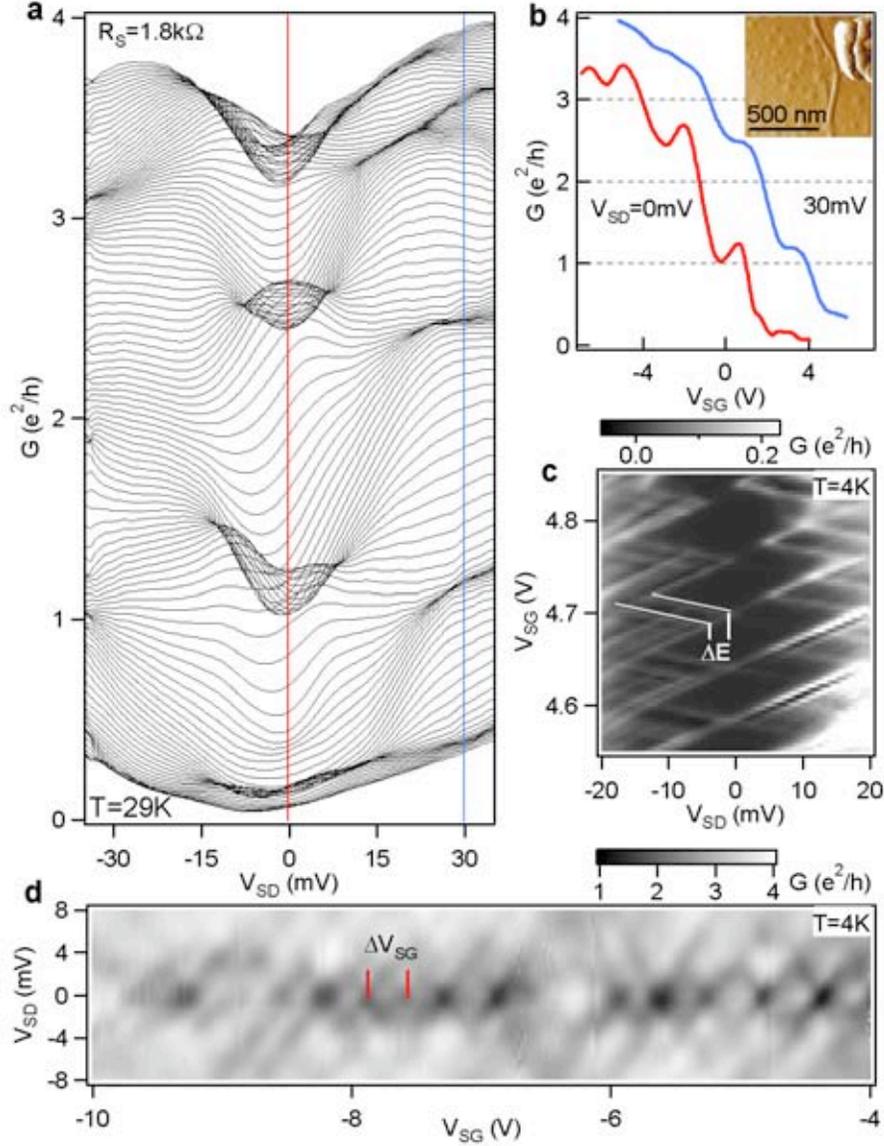

FIG. 1: (a) Differential conductance $G = dI/dV$ as a function of source drain bias, $V_{SD}$, and side gate voltage, $V_{SG}$, for a bent nanotube device, diameter $d \sim 3.5$ nm, at temperature $T = 29$ K. Series resistance $R_s$ is indicated. Traces are taken at fixed $V_{SG}$; bunched traces correspond to conductance plateaus. (b) Slices from (a) at fixed $V_{SD}$ as a function of $V_{SG}$. The high-bias trace is offset by 2V in $V_{SG}$ for clarity. Inset: AFM image of the device, showing a tube pushed toward the side gate (top right). Total device length is $\sim 1.5 \mu$m. (c) $G$ measured at 4 K for the same device with $V_{SG}$ set below the first plateau where Coulomb blockade diamonds are evident. Typical excited state level spacings, $\Delta E$, are 2-3 meV, corresponding to a device length L $\sim$ 500-700 nm. (d) $G$ in the high conductance region ($G > e^2/h$), with Fabry-Perot interference period $\Delta V_{SG} \sim 0.3$ V. For this panel, series resistance $R_S$=1.2 k$\Omega$.

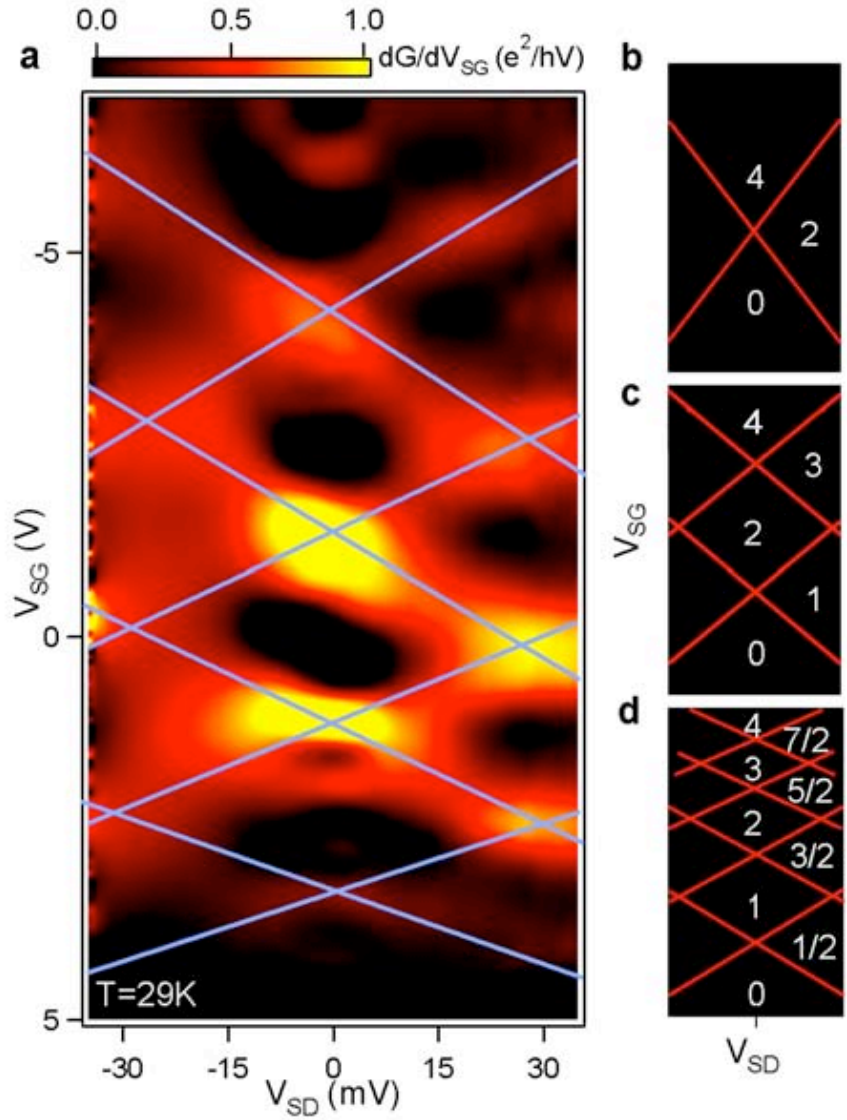

FIG. 2: (a) Transconductance, $dG/dV_{SG}$, as a function of $V_{SD}$ and $V_{SG}$, for data in Fig. 1a. Dark regions correspond to plateaus, bright regions to transitions between plateaus. Blue lines are guides indicating the evolution of conductance modes with $V_{SD}$ and $V_{SG}$. (b-d) Models for transconductance evolution with gate voltage for a four-mode 1D conductor for the case of (b) four-fold degeneracy, (c) two-fold degeneracy, (d) fully broken degeneracy. Numbers denote expected conductance values for plateaus, in units of $e^2/h$ (see text).

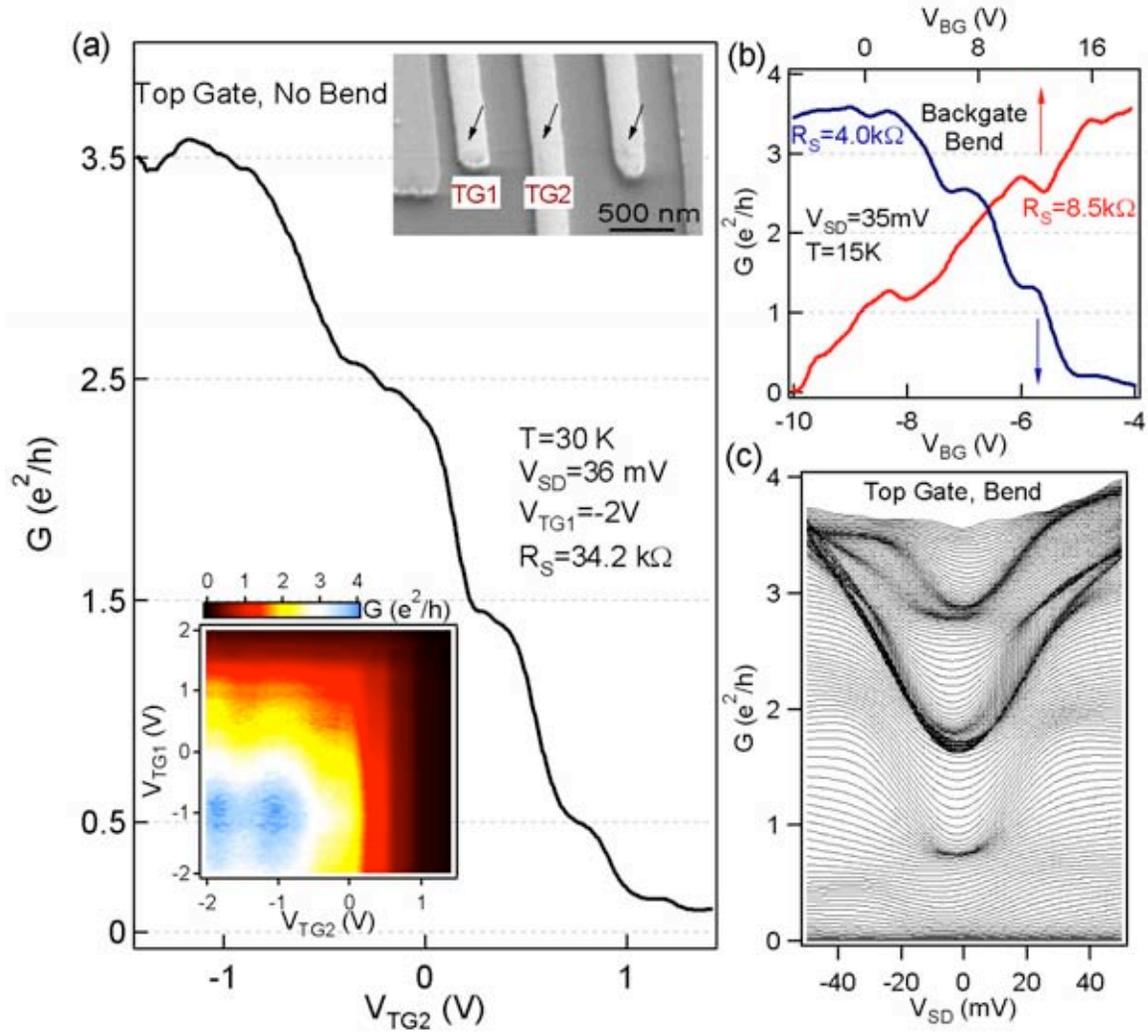

FIG. 3: (a) Differential conductance $G$ as a function of gate voltage for an unbent nanotube, diameter $d \sim 1.5$ nm, with top gates, showing plateaus spaced by $\sim e^2/h$. Upper Inset: SEM of measured device. The nanotube (arrows) is visible under the $SiO_2$ and top gates. Lower Inset: $G$ as a function of two top gates, with $V_{SD} = 36$ mV. (b) $G(V_{BG})$ at $V_{SD} = 35$ mV for another device, with $d \sim 3.2$ nm. Conductance plateaus appear for both hole (blue trace, bottom axis) and electron (red trace, top axis) transport. Different series resistances have been subtracted from the traces as noted on the figure. (c) $G$ as a function of $V_{SD}$ and top gate voltage at $T = 50$ K for a bent-tube device, $d \sim 4.5$ nm, with a top-gate. $R_S = 11.8$ k$\Omega$.

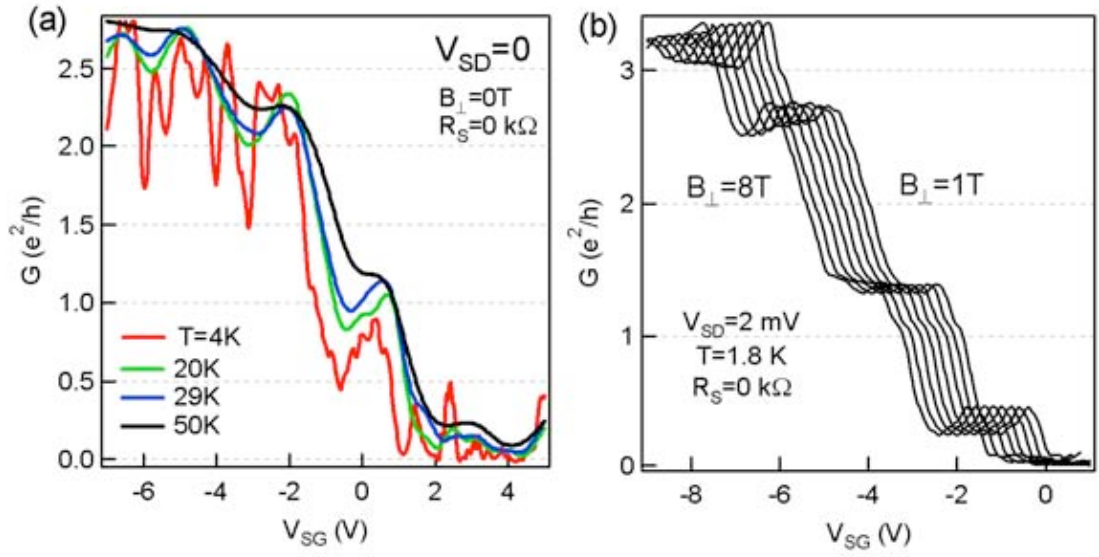

FIG. 4 (a) Temperature dependence of conductance plateaus around $V_{SD} = 0$, for the device in Fig. 1b (inset). (b) Evolution of the low-bias conductance plateaus for the same device (different cooldown) in a magnetic field perpendicular to the sample plane with $V_{BG} = 0.68$ V. Curves are offset in $V_{SG}$ for clarity. No series resistance is subtracted in (a) or (b).